\documentclass[prl,showpacs,twocolumn,preprintnumbers,amsmath,amssymb,superscriptaddress]{revtex4-1}

\usepackage{graphicx}
\usepackage{dcolumn}
\usepackage{bm}
\usepackage[latin1]{inputenc}
\usepackage{amssymb}
\usepackage{amsthm} 
\usepackage{multirow}
\usepackage{color}
\usepackage{dsfont}
\usepackage{blindtext}

\newcommand{\ud}[1]{{#1^{\dagger}}}
\newcommand{\bra}[1]{\left\langle #1\right|}
\newcommand{\ket}[1]{\left| #1\right\rangle}

\newcommand\Tr{\mathrm{Tr}}
\newcommand{\mean}[1]{\langle#1\rangle}

\begin{document}


\title{Spontaneous, collective coherence in driven, dissipative cavity
  arrays}

\author{J. Ruiz-Rivas}
\affiliation{Departament d'\`Optica, Universitat de Val\`encia, Dr. Moliner 50, 46100 Burjassot, Spain}
\author{E. del Valle}  \email[]{elena.delvalle.reboul@gmail.com}
\affiliation{F\'isica Te\'orica de la Materia Condensada, Universidad Aut\'onoma de Madrid, 28049 Madrid, Spain}
\author{C. Gies}
\affiliation{Institute for Theoretical Physics, University of Bremen, 28334 Bremen, Germany}
\author{P. Gartner}
\affiliation{Institute of Physics and Technology of Materials, P.O. Box MG-7, Bucharest-Magurele, Romania}
\author{M. J. Hartmann}
\affiliation{Institute of Photonics and Quantum Sciences, Heriot-Watt University, Edinburgh, EH14 4AS, United Kingdom}
\affiliation{Technische Universit\"at M\"unchen, Physik Department, James Franck Str., 85748 Garching, Germany}

\date{\today}

\begin{abstract}
  We study an array of dissipative tunnel-coupled cavities, each
  interacting with an incoherently pumped two-level emitter.  For
  cavities in the lasing regime, we find correlations between the
  light fields of distant cavities, despite the dissipation and the
  incoherent nature of the pumping mechanism.  These correlations
  decay exponentially with distance for arrays in any dimension but
  become increasingly long ranged with increasing photon tunneling
  between adjacent cavities. The interaction-dominated and the
  tunneling-dominated regimes show markedly different scaling of the
  correlation length which always remains finite due to the finite
  photon trapping time. We propose a series of observables to
  characterize the spontaneous build-up of collective coherence in the
  system.
\end{abstract}

\pacs{67.25.dj,42.50.Ct,64.60.Ht,42.55.Ah}

\maketitle

Arrays of optical or microwave cavities, each interacting strongly
with quantum emitters and mutually coupled via the exchange of
photons, have been introduced as prototype setups for the study of
quantum many-body physics of
light~\cite{hartmann2006,greentree2006,angelakis2007}. Even though
ground or thermal equilibrium states of the corresponding quantum
many-body systems are challenging to generate in experiments, much of
the initial attention has focussed on this
regime~\cite{hartmann2008,tomadin2010,houck2012,Carusotto13}. In any
realistic experiment with cavity arrays, however, photons are
dissipated due to the imperfect confinement of the light, and emitter
excitations have finite lifetimes.  It is thus crucial and useful to
explore the driven-dissipative regime of these structures, where
photon losses are continuously compensated by pumping new photons into
the cavities. A special role is here taken by the stationary states
where photon pumping and losses balance each other in a
\emph{dynamical equilibrium}. This regime has thus received
considerable attention in recent
years, where coherent and strongly correlated phases have been
discovered \cite{carusotto2009,Hartmann10,nissen2012}, but also
analogies to quantum Hall physics \cite{Umucalilar12} and
topologically protected quantum states \cite{Bardyn12} have been
discussed.

In previous investigations of coupled cavity arrays in
driven-dissipative regimes, the pump mechanism that injects photons
into the array has been assumed to be a coherent drive at each cavity
\cite{carusotto2009,Hartmann10,nissen2012,Umucalilar12,Bardyn12}.
Therefore any phase-coherence between light fields in distant cavities
that was seen in these studies can at least in part be attributed
to the fixed phase relation between their coherent input drives.
Here, in contrast, we show that such a coherence between distant
cavities can build up spontaneously, triggered only by physical
processes within the array.  In this way we address the question of
whether a non-equilibrium superfluid can develop in these structures.
To this end, we consider a cavity array that is only driven by an
incoherent pump which explicitly avoids any external source for a
preferred phase relation between photons in different cavities.

In our model, each cavity strongly interacts with a two-level emitter.
Whereas both, emitters and cavity photons, are subject to dissipation
processes, the cavities are excited via the emitters only, which are
population inverted by an incoherent pump. For a single cavity our
model reduces to the previously considered and realized
\emph{one-emitter
  laser}~\cite{mu92a,mckeever03a,astafiev07a,nomura10a,delvalle11a}.
Generalizations of this single cavity model have also been studied for
two~\cite{yeoman98a} and multiple
emitters~\cite{laussy11a,auffeves11a,poddubny10a} or emitters
supporting multi-exciton states \cite{Gies:11}.

We focus our analysis on the build-up of first-order coherence between
the fields in distant cavities as this quantity is typically
considered for investigating long range order and the emergence of
superfluidity, e.g. in optical lattices \cite{BDZ07}.  In cavity
arrays these correlations can be measured by recording the
interference pattern of the light fields emitted from the individual
cavities.  We find that collective correlations indeed build up in our
set-up when the cavities are in the lasing regime.  These correlations
decay exponentially as the distance between the considered cavities
tends to infinity for any dimension of the array. As intuitively
expected, the associated correlation length increases with increasing
photon tunneling between the cavities. For the interaction-dominated
regime this increase is logarithmic, whereas it is a power law in the
tunneling-dominated regime. Nonetheless, for any non-vanishing cavity
decay rate, the correlation length always remains finite.

Related questions are of high relevance for ultra-cold
atoms~\cite{diehl2008}, ions~\cite{schindler13a} , superconducting
circuits~\cite{marcos12a} or
exciton-polariton condensates~\cite{Carusotto13}. For the latter,
functional renormalization group approaches showed that, correlations
at least decay exponentially in isotropic
two-dimensional~\cite{Altmann13} but can be long range in
three-dimensional systems~\cite{Sieberer13}.

Finally, we also find that the collective coherence build-up manifests
strongly in the local cavity properties such as intensity and spectrum
of emission. In particular, lasing and its typical photoluminescence (PL)
lineshape, the Mollow triplet~\cite{delvalle10d,delvalle11a}, can be
observed far out of resonance between emitter and cavity as a result
of the emergence of collective photonic modes.

Suitable experimental platforms for exploring our findings are
superconducting circuit~\cite{houck2012}, photonic
crystal~\cite{majumdar12c,Rundquist:13},
micro-pillar~\cite{Abbarchi13}, or waveguide coupled
cavities~\cite{Lepert11}.

\begin{figure}
  \centering
  \includegraphics[width=0.86\linewidth]{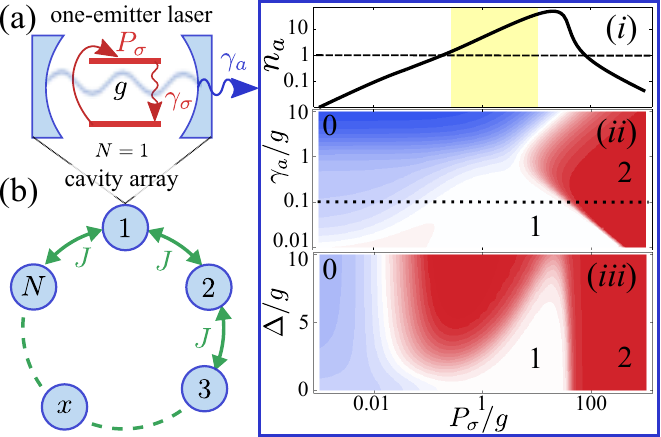}
  \caption{(Color online) (a) The building block of the array, the
    one-emitter laser and its main cavity emission properties: ($i$)
    cavity population $n_a$ as a function of $P_\sigma$ for
    $\gamma_a=0.1g$ and $\omega_{\sigma}=\omega_{a}$, with the lasing
    region highlighted in yellow.  Below, contour plots of $g^{(2)}$
    as a function of $P_\sigma$ and ($ii$) $\gamma_a$ at
    $\omega_{\sigma}=\omega_{a}$, or ($iii$)
    $\Delta=\omega_{\sigma}-\omega_{a}$ at $\gamma_a=0.1g$,
    with~$g^{(2)}>1$ in red, $g^{(2)}=1$ in white and $g^{(2)}<1$ in
    blue.  Also $\gamma_\sigma=0.01 g$ and~$J=0$.  (b) Scheme of the
    total system in one dimension: a circular array of~$N$ coupled
    cavities containing single emitters.}
  \label{fig:1}
\end{figure}

\emph{Model.}---We consider an array of cavities, each of which
interacts with a two level emitter, and is connected to adjacent
cavities via photon tunneling.  Our system, c.f. Fig.~\ref{fig:1}(a) and (b),
is thus described by a Jaynes-Cummings-Hubbard Hamiltonian
($\hbar=1$),
%
\begin{equation}\label{eq:hamiltonian}
  H=\sum_{j} H_{j}^{JC} + \sum_{<j,l>} J  [\ud{a_j}a_{l}+\ud{a_{l}}a_j] 
\end{equation}
with $H_{j}^{JC} = \omega_a\ud{a_j}a_j+\omega_\sigma\ud{\sigma_j}\sigma_j+g(\ud{a_j}\sigma_j+a_j\ud{\sigma_j})$,
where~$a_j$ is the photon annihilation operator and
$\sigma_j=\ket{\mathrm{g}}_j\bra{\mathrm{e}}_j$ the emitter
de-excitation operator in cavity~$j$. We assume periodic boundary
conditions and a homogeneous array with photon tunneling rate~$J$ so
that all~$H_{j}^{JC}$ feature the same photon frequency~$\omega_a$,
emitter transition frequency~$\omega_\sigma$, and light-matter
coupling $g$. We are interested in a driven-dissipative regime, where
each emitter is excited by an incoherent pump at a rate
$P_\sigma$~\cite{delvalle09a}, and decays spontaneously at a
rate~$\gamma_\sigma$.  The cavity photons in turn are lost at a
rate~$\gamma_{a}$ from each cavity. The dynamics of our system,
including these incoherent processes, follows the master equation,
$ \partial_t\rho=-i[H,\rho]+\sum_{j}
[\gamma_a\mathcal{L}_{a_j}+\gamma_\sigma\mathcal{L}_{\sigma_j}+P_\sigma\mathcal{L}_{\sigma_j^\dagger}](\rho)$,
where $\rho$ is the density matrix of the total system and
$\mathcal{L}_c(\rho)=\frac{1}{2}(2c\rho\ud{c}-\ud{c}c\rho-\rho\ud{c}c)$. We
are interested in the steady state ($\partial_t \rho = 0$) and neglect
pure dephasing, since it does not modify the results apart from
increasing the decoherence that~$P_\sigma$ already induces.

It is useful to introduce Bloch modes for the photons~\cite{hartmann10a} to
diagonalize the cavity part of Hamiltonian~(\ref{eq:hamiltonian}). For
a rectangular lattice of cavities of dimension $m$ and edge length
$N$, these modes read $p_{\vec{k}}= N^{-m/2}\sum_{\vec{r}}
e^{i\vec{k}\cdot \vec{r}} a_{\vec{r}}$, where $\vec{r}$ is an
$m$-dimensional lattice site index and the
Hamiltonian~(\ref{eq:hamiltonian}) takes the form $ H= \sum_{\vec{k}}
\omega_{\vec{k}}\ud{p_{\vec{k}}}p_{\vec{k}}+
\sum_{\vec{r}}\omega_\sigma\ud{\sigma_{\vec{r}}}\sigma_{\vec{r}}+\sum_{\vec{k},\vec{r}}(G_{\vec{k}\vec{r}}
p_{\vec{k}}\ud{\sigma_{\vec{r}}} +\text{h.c.})$, with
$\omega_{\vec{k}}=\omega_a+2J\sum_{\alpha = 1}^{m} \cos k_{\alpha}$,
$G_{\vec{k}\vec{r}} = g N^{-m/2} e^{-i\vec{k}\cdot \vec{r}}$, and
$k_{\alpha}=\frac{2\pi}{N}[-N/2+l_{\alpha}]$ for $N$ even or
$k_{\alpha}=\frac{2\pi}{N}[-(N+1)/2+l_{\alpha}]$ for $N$ odd
($l_{\alpha}=1,\ldots,N$).  The Bloch modes form a band with their
frequencies $\omega_{\vec{k}}$ distributed across the interval
$[\omega_{a} -2mJ, \omega_{a} +2mJ]$.  As easily seen, all modes
$p_{\vec{k}}$ decay at the same rate $\gamma_{a}$.  Hence, we have
mapped our model to a set of independent harmonic modes that all
couple to the same set of emitters with complex coupling constants
$G_{\vec{k}\vec{r}}$.  It is useful to define for each mode, the
detuning $\Delta_{\vec{k}}=\omega_\sigma-\omega_{\vec{k}}$, the total
decoherence rate $\Gamma=\gamma_a+P_\sigma+\gamma_\sigma$, the
effective coupling
$g_{\vec{k}}^\mathrm{eff}=g/\sqrt{1+(2\Delta_{\vec{k}}/\Gamma)^2}$,
and the population transfer from the emitters to the mode
(\emph{Purcell rate}) $F_{\vec{k}}=4(g_{\vec{k}}^\mathrm{eff})^2 /
\Gamma$.  Each Bloch mode can thus be driven by coherent excitation
exchange with the $N$ emitters.

Before analyzing the entire array we briefly review the properties of
a single site, the one-emitter laser, which provides a guideline for
our approach. In Fig.~\ref{fig:1}(a) we show the population,
$n_{a}=\mean{a^\dagger a}$, and second-order coherence function of a
single cavity, $g^{(2)}=\mean{a^\dagger a^\dagger a a}/\mean{a^\dagger
  a}$ as a function of~$P_\sigma$. In the strong coupling regime
($\gamma_a$, $\gamma_\sigma\ll g$) where we carry out our
investigations, one distinguishes~\cite{delvalle11a}: the linear and
quantum regimes at low pump
($g^{(2)}<1$)~\cite{averkiev09a,laussy11a,auffeves11a}, the lasing
regime ($g^{(2)}=1$), and the self-quenching and thermal regimes at
high pump ($1<g^{(2)}\leq 2$). In this work, we focus on the lasing
regime, where the emitter population is half-inverted,
$n_\sigma=\mean{\sigma^\dagger \sigma}\approx n_\sigma^\mathrm{L}
=1/2$, and the cavity accumulates a large number of photons,
$n_{a}\approx n_{a}^\mathrm{L}= P_\sigma/2\gamma_a$~\footnote{This is
  only below the maximum cavity population, reached at
  $P_\sigma\approx\kappa_\sigma/2$~\cite{delvalle11a}. We choose
  $\gamma_a=0.1g$, $\gamma_\sigma=0.01g$ and $P_\sigma=5g$ as a
  paradigmatic example of the lasing regime for any $N$.}.  Due to the
stochastic nature of the pump, $\mean{a}=0$~\cite{Molmer97}, and our
system can not be described by standard laser
theory~\cite{hanken_book84a}.  Instead, for the quantized light field,
photon-assisted polarizations $\mean{a^\dagger \sigma}$ are driven
\cite{gies07a} and induce the build-up of coherence in the cavity
field, for which $\mean{a^\dagger a \sigma^\dagger \sigma}\approx n_a
n_\sigma$.
These properties allow us to obtain simple rate equations for the
populations and polarizations that provide accurate results above the
quantum regime, i.e. for $P_\sigma>\gamma_a$,
$\gamma_\sigma$~\cite{delvalle11a}. The accuracy of this approach has
also been confirmed for $N>1$ emitters in a single
cavity~\cite{array_moelbjerg}.

\emph{Rate Equations.}---From the above master equation, we derive a hierarchy of coupled equations of motion for
correlators~\cite{supplemental} starting with
$n_{\sigma}=\mean{\sigma_{\vec{r}}^\dagger \sigma_{\vec{r}}}$ and
$n_{\vec{k}}=\mean{p_{\vec{k}}^\dagger p_{\vec{k}}}$. We apply
the cluster-expansion method up to order two~\cite{gies07a} to
truncate the equations. For the lasing and thermal regimes, this
approximation can be expected to be very accurate, thanks to the weak
and indirect interactions between modes or emitters, and it further allows us to assume
$\mean{\sigma_{\vec{r}}^\dagger \sigma_{\vec{s}}}\approx n_{\sigma}
\delta_{\vec{r},\vec{s}}$ and $\mean{p_{\vec{k}}^\dagger p_{\vec{q}}
  \sigma_{\vec{r}}^\dagger \sigma_{\vec{r}}}\approx n_{\vec{k}}
n_\sigma \delta_{\vec{k},\vec{q}}$ (indexes $\vec{r}$ and $\vec{s}$
label emitters and $\vec{k}$ and $\vec{q}$ label Bloch modes). We have numerically verified the validity of this approximation by including correlations between emitters in distant cavities. 
For the steady state we find
\begin{subequations}
  \label{eq:allrates}
  \begin{align} 
    \label{eq:rate-eqs-main-a}
    0 &=-\gamma_a n_{\vec{k}} +F_{\vec{k}} n_{\vec{k}} (2n_\sigma -1)+F_{\vec{k}}
    n_\sigma, \\
    0 & =P_\sigma -(P_\sigma+\gamma_\sigma +F) n_\sigma-(2n_\sigma-1) \tilde{F},
    \label{eq:rate-eqs-main-b}
\end{align}
\end{subequations}
with $F=N^{-m} \sum_{\vec{k}}F_{\vec{k}}$ and $\tilde{F} =
N^{-m}\sum_{\vec{k}} F_{\vec{k}} n_{\vec{k}}$. The polarizations are
then given by $\mean{p_{\vec{k}}^\dagger
  \sigma_{\vec{r}}}=iG_{{\vec{k}}{\vec{r}}}(n_\sigma-n_{\vec{k}}+2
n_{\vec{k}} n_\sigma)/(\Gamma/2+i\Delta_{\vec{k}})$ and the local
cavity populations by $n_a= N^{-m}\sum_{\vec{k}} n_{\vec{k}}$.
Eq. (\ref{eq:rate-eqs-main-a}) can be solved for
$n_{\vec{k}}$ to find
\begin{equation} 
  \label{eq:momentumdist}
  n_{\vec{k}} = \frac{\kappa_\sigma\Gamma}{4} \frac{n_{\sigma}}{(\delta/2)^{2} + \Delta_{\vec{k}}^{2}}
\end{equation}
with $\delta^{2} =\kappa_\sigma \Gamma \left[\Gamma/\kappa_\sigma- (2
  n_{\sigma} - 1) \right]$ and $\kappa_\sigma =4g^2/\gamma_a$, the
Purcell enhanced decay of an emitter through its local
cavity~\cite{delvalle11a}.  The distribution of Bloch mode populations
is thus a Lorentzian in~$\Delta_{\vec{k}}$ with width~$\delta$.

The central quantity of interest in our investigation are the
normalized correlations between cavity fields in distant
cavities~\cite{supplemental},
\begin{equation} \label{eq:correl} \mathcal{C}(\vec{r}) =
  \frac{\mean{a_{\vec{0}}^\dagger
      a_{\vec{0}+\vec{r}}}}{\mean{a_{\vec{0}}^\dagger a_{\vec{0}}}}
  =\frac{1}{n_{a} \, N^{m}}\sum_{\vec{k}}e^{-i \vec{k}\cdot\vec{r}}
  n_{\vec{k}},
\end{equation}
%
the Fourier transform of the Bloch mode populations $n_{\vec{k}}$.

\emph{Asymptotics of Correlations.}---Inserting
Eq.~(\ref{eq:momentumdist}) into Eq.~(\ref{eq:correl}), we find as a
central result that the correlations $\mathcal{C}(\vec{r})$ decay
faster than $r^{-n}$ as $r \to \infty$, where $r = |\vec{r}|$, for any
positive integer~$n$ and lattice dimension~$m$, provided $\delta \ne
0$.  The proof of this statement is provided in~\cite{supplemental},
and proceeds by showing, via multiple applications of the divergence
theorem, that $\sum_{\vec{r}} r^{2n} |\mathcal{C}(\vec{r})|^{2}$ is
finite for any positive integer $n$. The only possibility for the
system to become critical, in the sense that the correlation length of
$|\mathcal{C}(\vec{r})|$ diverges, would be that $\delta$ vanishes,
i.e. that $\Gamma/\kappa_\sigma = (2 n_{\sigma} - 1) $.  It is however
easily seen that the last term in Eq. (\ref{eq:rate-eqs-main-b})
diverges for $N\to\infty$ unless $(2 n_{\sigma} - 1) \to 0$, which,
for $\delta = 0$, would imply $\gamma_{a} = 0$. We, therefore, conclude
that any non-vanishing photon decay rate keeps the correlation length
finite and thus prevents criticality.

\begin{figure}
  \centering
  \includegraphics[width=\linewidth]{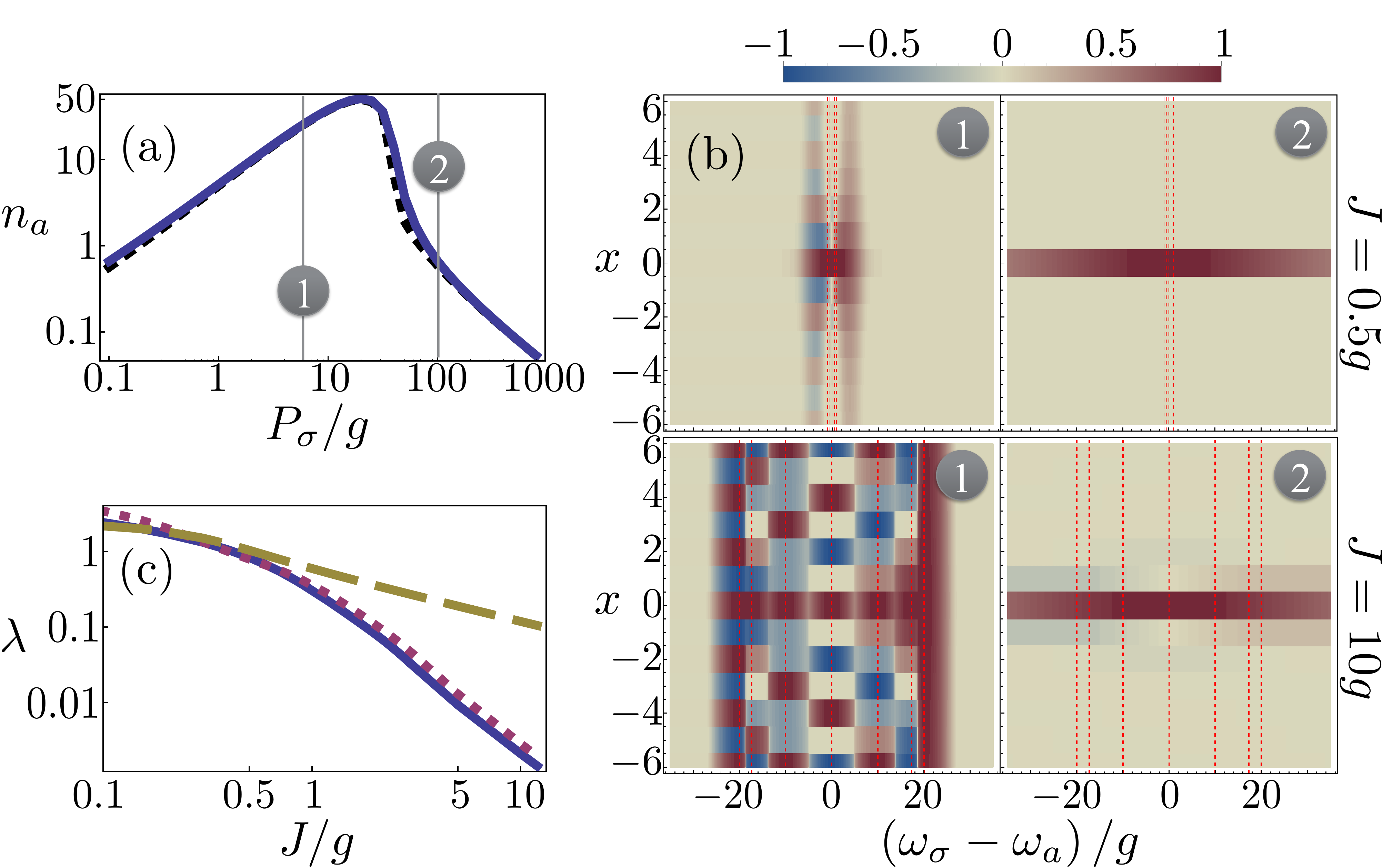}
  \caption{(a)~Cavity population~$n_a$ for~$\omega_{\sigma} =
    \omega_{a}$ as a function of pump~$P_{\sigma}$ for $J=0.5g$ (solid
    blue) and~$J=10g$ (dashed black), with~$N=12$, $\gamma_a=0.1g$,
    $\gamma_\sigma=0.01g$. (b)~Corresponding first order
    correlations~$\mathcal{C}(x)$ as a function of distance $x$ and
    emitter frequency $\omega_\sigma$ at pump rates~(1) and (2) in
    plot (a). Bloch mode resonances are plotted as vertical dashed red
    lines. (c)~Inverse correlation lengths, $\lambda$, as obtained
    from fits (see main text) for~$N=108$, $P_\sigma= 5 g$, and
    $\Delta = 0$ (solid), $\Delta = J$ (dotted) or $\Delta = 2J$
    (dashed).}
  \label{fig:3}
\end{figure}

\emph{Correlations in one dimension (1D).}---We now examine
correlations in a 1D chain, $\mathcal{C}(x)$ with $-N/2\leq x \leq
N/2$, Eq.~(\ref{eq:correl}),
considering $N$ to be a multiple of 4, so that
the Bloch modes are distributed symmetrically around the cavity
frequency.
We first focus on~$N=12$ with~$J=0.5g$ or $10g$, for which we
show~$n_a$ as a function of the pump in Fig.~\ref{fig:3}(a). Both
cases undergo very similar and characteristic transitions into and
out of lasing (c.f. Fig.~\ref{fig:1}($i$)). We select two
pumping rates representative of the lasing~(1) and thermal~(2) regimes
and plot~$\mathcal{C}(x)$ as a
function of detuning~$\Delta=\omega_\sigma-\omega_a$ and the
separation~$x$ between the cavities in Fig.~\ref{fig:3}(b).
For~$|\Delta| < 2 J$, $\mathcal{C}(x)$
oscillates as $\cos(\overline{k} x)$, where~$\overline{k}$
and~$-\overline{k}$ are the (degenerate) modes closest to resonance
with the emitters, i.e. $|\Delta| \approx 2 J \cos \overline{k}$.
The correlation length is longer in the
lasing regime~(1), increases for larger~$J$ and becomes maximal
for~$|\Delta| = 2 J$ in each case, i.e. when the emitters are in
resonance with the edges of the Bloch band.  For~$J=10g$ it becomes
larger than the finite size array of~$N = 12$ considered here since
the frequency separation between Bloch modes is so large that the
emitters only populate one mode efficiently.  Note that any decay of
correlations is entirely due to destructive interference between
different Bloch-mode contributions.

Let us now explore~$|\Delta| \le 2J$, where the emitters are on
resonance with the Bloch band and photonic modes are appreciably
populated.  For a long chain, $N \gg 1$, and large tunneling rates, $J
\gg g$, analytical estimates can be found for the correlations
$\mathcal{C}(x)$~\cite{supplemental}. In agreement with
Fig. \ref{fig:3}, these show exponential decay modulated by an
oscillation.  We thus fit a function $f(x) = [c_{1} \cos(\nu x) +
c_{2} \sin(\nu x)] \exp(-\lambda x)$ to $\mathcal{C}(x)$ in the entire
range of tunneling rates $J$ and extract the inverse correlation
length, $\lambda$, from the fit (see \cite{supplemental} for
examples).  Fig. \ref{fig:3}(c) shows $\lambda$ for three cases:
$\Delta = 0$ (solid), $\Delta = J$ (dotted) and $\Delta = 2J$ (dashed)
for a chain of $N=108$ cavities, which has Bloch modes in resonance
with the emitters for all considered values of $\Delta$ so that
finite-size effects are suppressed.  As second main result of our work
we observe a clear transition from the regime with $J < g$, where
$\lambda \propto - \ln J$, to the regime $J > g$, where $\lambda
\propto J^{-1}$ for $J\gg |\Delta|$ and $\lambda \propto J^{-1/2}$ for
$2 J = |\Delta|$~\cite{supplemental}.  These behaviors are also found
from analytical estimates for $N\to \infty$~\cite{supplemental}.

\begin{figure}
  \includegraphics[width=\linewidth]{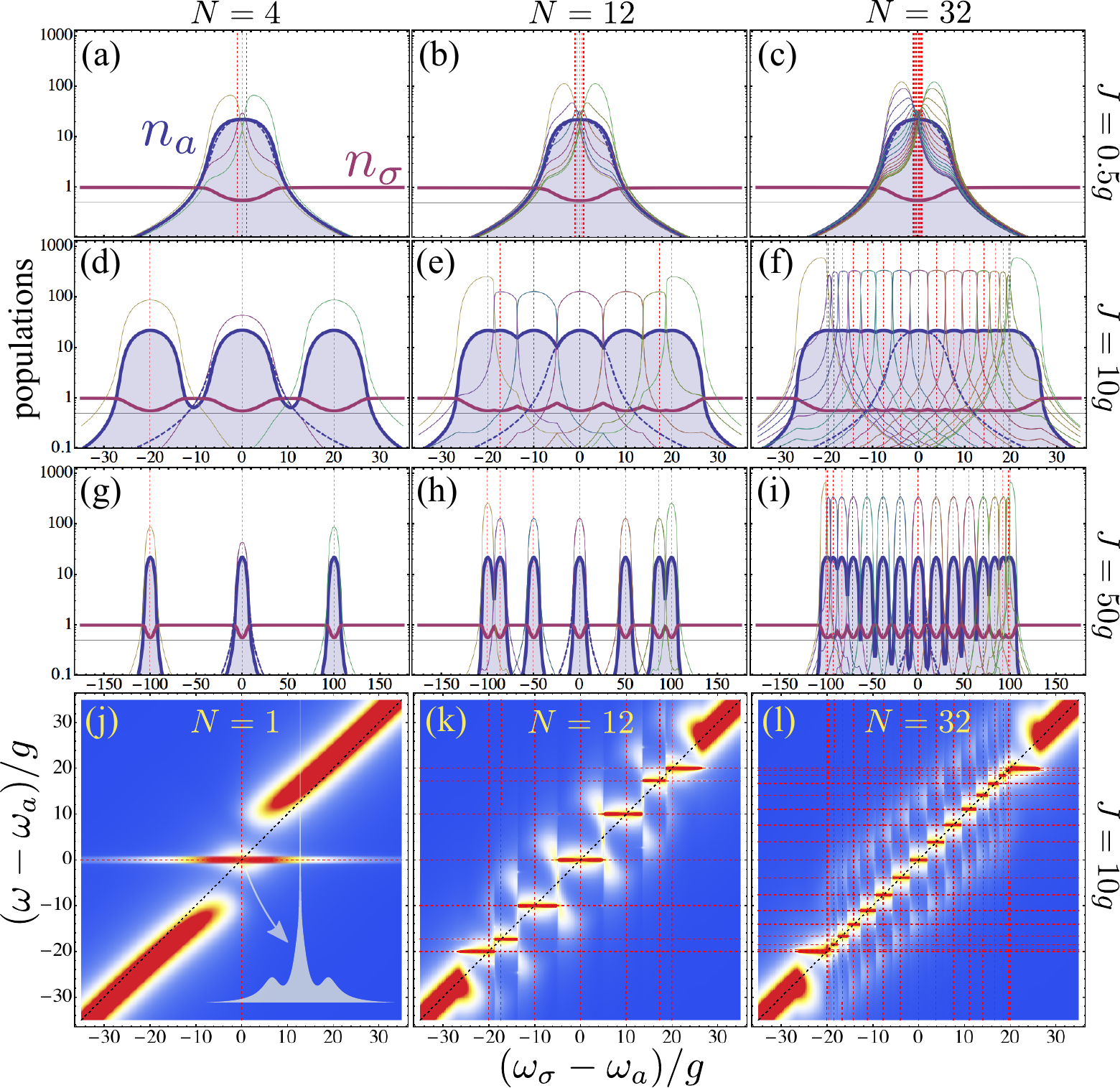}
  \caption{(a)--(i) Populations of the different modes involved, when
    sweeping the emitter frequency $\omega_\sigma$ through the system
    resonances (vertical red dashed lines): $n_a$ in solid and filled
    blue, $n_\sigma$ in solid pink, the Bloch modes $n_k$ with thin
    lines and $n_a$ for the case $N=1$ in dashed blue as a
    reference. (j) Emitter spectrum of emission for~$N=1$ and varying
    $\omega_\sigma$, showing a Mollow triplet around resonance. In
    inset, the lineshape at resonance. In (k) and (l), the spectra for
    cases (e) and (f), respectively. We use a temperature color code
    which goes from blue (0) to red (maximum values). Parameters are
    $N=4$, 12, 32 and $J=0.5g$, $10g$, $50g$, varying as
    indicated. Also: $P_\sigma=5g$, $\gamma_a=0.1g$,
    $\gamma_\sigma=0.01g$.}
 \label{fig:2}
\end{figure}

\emph{Local properties in 1D chains.}---Finally, we present some
experimentally observable and distinctive local signatures of the
collective lasing regime in the array, as a function of $\Delta$. In
Fig.~\ref{fig:2}(a)--(i) we plot~$n_a$ and~$n_\sigma$, computed from
Eqs.~(\ref{eq:allrates}), for various arrays. Each underlying Bloch
mode~$n_k$ enters its own lasing regime at~$\omega_\sigma=\omega_k$.
This results in the enhancement of~$n_a$ to a fixed value, given by
the resonant one-emitter case~$n_a^\mathrm{L}$, while the emitter
population decreases to $n_\sigma^\mathrm{L}\approx 1/2$ from its
saturation value of 1.
Note that these traits are independent of $g$, $N$ and $J$ once the
system is strongly enough coupled to reach the lasing
regime~\cite{laussy12d}.
Interactions as small as $J\lesssim 0.5g$ (Fig.~\ref{fig:2} upper row)
are not enough to make a qualitative difference from the $N=1$ case in
the local populations.
The width in detuning of the apparent single broad resonance is given
by
$2\Delta_\mathrm{max}=\sqrt{P_\sigma(\kappa_\sigma-P_\sigma)}$~\footnote{Estimation
  obtained by solving $n_a\approx
  n_a^\mathrm{L}[1-\frac{P_\sigma}{\kappa_\sigma}(1+(\frac{2\Delta}{P_\sigma})^2)]=0$
  in the detuned one-emitter laser~\cite{delvalle11a}.}.
Increasing interactions, $J>g$ (other rows), splits the Bloch modes
apart so that they can be selectively addressed by changing
detuning. The excitation is distributed equally among the driven modes
so, at resonance, $n_{k=0,\pi}=N n_a^\mathrm{L}$ and $n_{\pm k}=N
n_a^\mathrm{L}/2$ for the other central modes. This results in a
series of peaks for $n_a$ of equal height~$n_a^\mathrm{L}$ and
width~$2\Delta_\mathrm{max}$. When the width is smaller than the
average separation between Bloch modes, approximately given by $4J/N$
(or $4J/(N-1)$ for odd $N$), a plateau forms in the populations that
extends for~$|\Delta| \leq 2J$, c.f.  Fig.~\ref{fig:2}(f). At this
point, increasing~$N$ does not affect the results qualitatively.

Another very distinctive feature of the collective lasing is provided
by the PL spectrum. Despite the incoherent pump, a Mollow triplet
forms~\cite{mollow69a,delvalle10d,delvalle11a,delvalle13c_mathematica}
whenever $\omega_\sigma=\omega_k$ for some $k$, thanks to the
effective multi-Bloch-mode coherent drive~$\Omega(t)=\sum_k g
\sqrt{n_k/N} e^{-i\omega_k t}$~\cite{supplemental}.  In
Fig.~\ref{fig:2}(j)--(l), we compare~$N=1$, 12 and 32, for
varying~$\Delta$.  The Rayleigh peak, pinned at the laser frequency
for a single mode excitation~\cite{mollow69a}, jumps from Bloch mode
to Bloch mode, depending on which one dominates, in correspondence
with the population plateaus of Fig.~\ref{fig:2}(e),~(f). The
sidebands are positioned at~$\omega_k\pm
2\sqrt{2}g\sqrt{n_a^\mathrm{L}}$, around resonance with a degenerate
Bloch mode~$\omega_k$, and at $\omega_k\pm 2g\sqrt{n_a^\mathrm{L}}$,
with the edge modes. Therefore, high $N$ and closely packed Bloch
modes
give rise to two Mollow continuous sidebands at $\omega_\sigma\pm
2\sqrt{2}g\sqrt{n_a^\mathrm{L}}$, extending over $|\Delta|\leq
2J$.

\emph{Acknowledgements.}---JR-R acknowledges the hospitality of
Technical University Munich, where part of this work was done.  EdV
acknowledges support from the Alexander von Humboldt-Foundation and
the Spanish MINECO under contract MAT2011-22997 and MJH from the Emmy
Noether grant HA 5593/1-1 and the CRC 631 (both DFG).

\bibliography{Sci,array,books}

\pagebreak
$ $
\appendix

\begin{center}
{\Large {\bf Supplemental Material}}
\end{center}

\setcounter{equation}{0}
\setcounter{figure}{0}

\section{I. Equations of motion for the correlators}
\label{ap:qrf}

In this section, we derive the system equations of motion in the case of a
one-dimensional array. They can be trivially extended to higher
dimensions.

The most general operator in the system reads
$\mean{O}=\mean{\Pi_{k}p_k^{\dagger m_k}p_k^{n_k}
  \Pi_{j}\sigma_1^{\dagger\mu_j}\sigma_1^{\nu_j}}$. From the master
equation in the main text, we obtain the equations of motion for the
set of relevant operators by means of the general relation $\partial_t
\mean{O}=\Tr(O \partial_t\rho)$ as
\begin{multline}
  \label{eq:TueMay5174356GMT2009}
  \partial_t \mean{\Pi_{k}p_k^{\dagger m_k}p_k^{n_k}
    \Pi_{j}\sigma_1^{\dagger\mu_j}\sigma_1^{\nu_j}}
  =\sum_{\bar m_1,\bar n_1,\ldots\bar\mu_1,\bar\nu_1\dots}\\
R_{ \tiny
\begin{array}{c}
m_1,n_1,\ldots\mu_1,\nu_1\dots\\
\bar m_1,\bar n_1,\ldots\bar\mu_1,\bar\nu_1\dots
\end{array}}
\mean{\Pi_{k}p_k^{\dagger \bar m_k}p_k^{\bar n_k} \Pi_{j}\sigma_1^{\dagger
    \bar \mu_j}\sigma_1^{\bar \nu_j}}\,.
\end{multline}
The diagonal elements in~$R$, involving all modes and emitters,
are given by~\cite{delvalle_book10a}:
\begin{eqnarray}
  \label{eq:TueDec23114907CET2008}
&&  R_{ \tiny
\begin{array}{c}
m_1,n_1,\ldots\mu_1,\nu_1\dots\\
m_1,n_1,\ldots\mu_1,\nu_1\dots
\end{array}}=\\
&&\sum_k [i\omega_k
(m_k-n_k)-\frac{\gamma_a}2(m_k+n_k)]\nonumber\\
&+&\sum_j[i\omega_\sigma(\mu_j-\nu_j)-\frac{\gamma_\sigma+P_\sigma}{2}(\mu_j+\nu_j) -\frac{\gamma_\phi}{2}(\mu_j-\nu_j)^2 ]\nonumber\,.
\end{eqnarray}
We have included in these elements the effect of pure dephasing at a
rate~$\gamma_\phi$, added to the master equations through the Lindblad
term $\gamma_\phi\mathcal{L}_{\sigma_j^\dagger \sigma_j} (\rho)$. This
only results in the increase of the total decoherence rate
into~$\Gamma=\gamma_a+P_\sigma+\gamma_\sigma+\gamma_\phi$~\cite{gonzaleztudela10b}. Next,
the incoherent pumping of emitter~$j$ affects only elements concerning
such emitter so that for all $j$:
\begin{equation}
  \label{eq:WedJun12112000CEST2013}
  R_{ \tiny
\begin{array}{c}
\ldots\mu_j,\nu_j\dots\\
\ldots\mu_j,\nu_j\dots
\end{array}}= P_\sigma \mu_j \nu_j\,.
\end{equation}
Finally, the coupling between mode $k$ and emitter $j$, provides the
elements:
\begin{subequations}
\begin{align}
  & R_{ \tiny
\begin{array}{c}
m_k,n_k,\mu_j,\nu_j\\
m_k-1,n_k,1-\mu_j,\nu_j 
\end{array}}=iG_{kj}m_k(1-\mu_j) \,, \\
  & R_{ \tiny
\begin{array}{c}
m_k,n_k,\mu_j,\nu_j\\
m_k,n_k-1,\mu_j,1-\nu_j 
\end{array}}=-iG_{kj}^*n_k(1-\nu_j) \,,\\
  & R_{ \tiny
\begin{array}{c}
m_k,n_k,\mu_j,\nu_j\\
m_k+1,n_k,1-\mu_j,\nu_j 
\end{array}}=iG_{kj}^*\mu_j \,, \\
  & R_{ \tiny
\begin{array}{c}
m_k,n_k,\mu_j,\nu_j\\
m_k,n_k+1,\mu_j,1-\nu_j 
\end{array}}=-iG_{kj}\nu_j\,\\
  &R_{ \tiny
\begin{array}{c}
m_k,n_k,\mu_j,\nu_j\\
m_k+1,n_k,\mu_j,1-\nu_j 
\end{array}}=-2iG_{kj}^* \mu_j(1-\nu_j) \,,\\
  & R_{ \tiny
\begin{array}{c}
m_k,n_k,\mu_j,\nu_j\\
m_k,n_k+1,1-\mu_j,\nu_j 
\end{array}}=2iG_{kj} \nu_j(1-\mu_j)  \,,
\end{align}
\end{subequations}
and zero everywhere else. 

With these general rules, we can write the equations for the main
correlators of interest, starting with the populations of the modes,
$n_k=\mean{p^\dagger_k p_k}$ and emitters $n_j=\mean{\sigma_j^\dagger
  \sigma_j}$:
\begin{subequations}
    \label{eq:ThuJun13121117CEST2013}
  \begin{align}
    &\partial_t n_j =-(P_\sigma+\gamma_\sigma ) n_j+P_\sigma -2\sum_k \Im[G_{kj}^* \mean{p_k^\dagger \sigma_j}]\,,\\
    &\partial_t n_k =-\gamma_a n_k+2\sum_j   \Im[G_{kj}^* \mean{p_k^\dagger \sigma_j}]\,,\\
    &\partial_t \mean{p_k^\dagger
      \sigma_j}=-[\frac{\Gamma}{2}
    +i(\omega_\sigma-\omega_k)]\mean{p_k^\dagger
      \sigma_j}\nonumber\\
    &+iG_{kj}[n_j-n_k+2\mean{p_k^\dagger p_k   \sigma_j^\dagger \sigma_j}]\nonumber\\
    & +\sum_{l\neq j} i G_{kl} \mean{\sigma_l^\dagger
      \sigma_j}+\sum_{q\neq k} (-i G_{ql}) \mean{p_k^\dagger p_q }\nonumber\\
    &+\sum_{q\neq k}2 i G_{qj} \mean{p_k^\dagger p_q \sigma_j^\dagger
      \sigma_j}\,.
\end{align}
\end{subequations}
%
The equations for the correlators that represent the indirect coupling
between different emitters or Bloch modes are:
\begin{subequations}
    \label{eq:FriJun14154756CEST2013}
  \begin{align}
    &\partial_t \mean{\sigma_l^\dagger
      \sigma_j}=-(P_\sigma+\gamma_\sigma)
    \mean{\sigma_l^\dagger \sigma_j}\nonumber\\
    &+\sum_k   i[G_{kl}^*\mean{p_k^\dagger   \sigma_j}-G_{kj}\mean{p_k\sigma_l^\dagger}]\nonumber\\
    &+\sum_k 2i[G_{kj}\mean{p_k\sigma_l^\dagger \sigma_j^\dagger
      \sigma_j}-G_{kl}^*\mean{p_k^\dagger \sigma_l^\dagger
      \sigma_l \sigma_j}]\,, \label{eq:sigsigcorr}\\
    &\partial_t \mean{p_k^\dagger
      p_q}=-[\gamma_a-i(\omega_k-\omega_q)]
    \mean{p_k^\dagger p_q}\nonumber\\
    &+\sum_j
    i[G_{kj}\mean{p_q\sigma_j^\dagger}-G^*_{qj}\mean{p_k^\dagger
      \sigma_j}]\,.\label{eq:SatAug17123213CEST2013}
  \end{align}
\end{subequations}
Within the formal scheme of the Cluster-Expansion method,
Eq.~\eqref{eq:sigsigcorr} is of the same order as the Bloch-mode
populations $n_k$. This is owed to the dominant Jaynes-Cummings
interaction in the system, which can be used to establish a formal
equivalence between an electronic transition and photon creation or
absorption~\cite{gies07a}. In the thermal and lasing regimes
investigated in the main text, the influence of these correlations is
small and, therefore, neglected in order to keep the formal solution
of the equations as simple as possible.

Finally, the intensity-intensity correlations are given by:
%
 \begin{align}
   &\partial_t \mean{p_k^\dagger p_k \sigma_l^\dagger\sigma_l}=-(\gamma_a+P_\sigma+\gamma_\sigma) \mean{p_k^\dagger p_k \sigma_l^\dagger\sigma_l}+P_{\sigma}n_k \nonumber\\
   &+i(G_{kl}^*\mean{p_k^\dagger p_k^\dagger p_k \sigma_l}-G_{kl}\mean{p_k^\dagger p_k p_k \sigma_l^\dagger}) \nonumber\\
   &+i\sum_{q\neq k}(G_{ql}^*\mean{p_q^\dagger p_k^\dagger p_k \sigma_l}-G_{ql}\mean{p_q^\dagger p_k p_k \sigma_l^\dagger}) \nonumber\\
   &+i\sum_{j\neq l}(G_{kj}\mean{p_k\sigma_l^\dagger \sigma_l
     \sigma_j^\dagger}-G_{kj}^*\mean{p_k^\dagger \sigma_j
     \sigma_l^\dagger \sigma_l })\,.
\end{align}
\label{eqs}
%
Thanks to the translational invariance in the array (which leads to
linear momentum conservation), the Bloch mode correlations vanish,
$\mean{p_{\vec{k}}^\dagger
  p_{\vec{q}}}=\delta_{\vec{k},\vec{q}}n_{\vec{k}}$, and the
cavity correlations are simply the Fourier transform of the Bloch mode
populations:
\begin{equation}
  \label{eq:FriJun14170914CEST2013}
  \mean{a_j^\dagger a_{j+x}}=\frac{1}{N}\sum_ke^{-ix k} n_k\,,
\end{equation}
and Eq.~(4) from the main text, more generally stated in any
dimension.

\subsubsection*{Analytical solutions of the rate equations for $N=1$}
\label{ap:qrf2}

In the case $N=1$, we have only a single emitter and photonic mode so
$F_k\rightarrow F$ and the rate equations in the steayd state reduce
to:
\begin{subequations}
  \label{eq:rate-eqs}
  \begin{align}
    &0=-\gamma_a n_a +F n_a (2n_\sigma -1)+F n_\sigma\,, \nonumber\\
    &0=-(P_\sigma+\gamma_\sigma +F) n_\sigma+P_\sigma -(2n_\sigma-1) F
    n_a\,.\nonumber
\end{align}
\end{subequations}
The solution of these equations reads,
%
\begin{eqnarray}
  \label{eq:MonOct28124833CET2013}
    n_a &=&\frac{F(2 P_\sigma-\zeta_\sigma -\gamma_a)-\gamma_a
      \zeta_\sigma+\chi^2}{4F\gamma_a}\, , \\
      n_\sigma &=& \frac{P_\sigma-\gamma_a n_a}{\zeta_\sigma}
\end{eqnarray}
%
with $\chi^2 = \sqrt{[F(2 P_\sigma+\zeta_\sigma+\gamma_a)+\gamma_a
  \zeta_\sigma]^2-8F P_\sigma \zeta_\sigma(F+\gamma_a)}$ and
$\zeta_{\sigma} = P_{\sigma} + \gamma_{\sigma}$.
        
\section{II. Fast decay of correlations}

Here we consider a rectangular $m$-dimensional lattice
of cavities in the thermodynamic limit, i.e. where infinitely many
cavities are arranged in each lattice direction. We thus have a
continuum of momentum modes and  $\frac{1}{N^m}\sum_{\vec{k}}$ turns into
an integral over the Brillouin Zone (BZ) $V_k$ formed by the $m$-dimensional
cube extending from $-\pi$ to $\pi$ in each direction.  

The field correlations are given by
\begin{equation}
  \mathcal{C}(\vec{r}) = \frac{\mean{a_{\vec{0}}^\dagger
a_{\vec{0}+\vec{r}}}}{\mean{a_{\vec{0}}^\dagger a_{\vec{0}}}}=
\frac{1}{n_{a}\,(2 \pi)^{m}} \int_{V_{k}} d^{m}k e^{-i \vec{k} \vec{r}}
n(\vec{k}), 
\label{eq:corr}
\end{equation}
with $\vec r$ running on the lattice of $m$-dimensional vectors with integer
coordinates. 

For $\delta^2 >0$, $n(\vec k)$ is a continuous function of $k$ defined on a
finite domain, and therefore it is integrable over $V_k$. In this case the
Riemann-Lebesgue lemma ~\cite{rudin} ensures that $\mathcal{C}(\vec{r})$
decays to zero for $\vec r \to \infty$. The result we want to show is that this
decay is actually faster than any power of $r$. The proof relies essentially on
the fact that $n(\vec k)$ depends on $\vec k$ through cosine functions of the
components of $\vec k$. As such, $n(\vec k)$ and all its derivatives are
continuous and periodic functions of $\vec k$. By periodicity here we mean
invariant with respect to translations by reciprocal lattice vectors, i.e. 
$n(\vec k) = n(\vec k + \vec K)$, where the coordinates of $\vec K$ are integer
multiples of $2\pi$. In particular, on the surface of the BZ one finds pairwise
opposite points, differing by a reciprocal lattice vector. It follows that in
such points $n(\vec k)$ has equal values, and the same is true for all its
derivatives. 

For the proof we denote by $\alpha = \{\alpha_1, \alpha_2 \dots \alpha_m\}$ a
multi-index of natural numbers and by $|\alpha|$ the sum of its components
$\alpha_1 + \dots \alpha_m$. We denote also by $r^\alpha$ the quantity 
$r_1^{\alpha_1} r_2^{\alpha_2}\dots r_m^{\alpha_m}$. 
The result we want to show is that for any $\alpha$ one has
$r^\alpha\mathcal{C}(\vec{r}) \to 0$ when $r \to \infty$. 

Indeed, multiplying the integral in Eq.~(\ref{eq:corr}) with $r^\alpha$ amounts
to applying the derivative operator $(i \partial)^\alpha = i^{|\alpha|}
\partial_1^{\alpha_1}\dots \partial_m^{\alpha_m}$ to the plane-wave factor
$e^{-i\vec k \vec r}$ under the integral. By $\partial_i$ we mean the
derivative with respect to $k_i$. All these derivatives can be transferred upon
$n(\vec k)$ by repeatedly applying the divergence theorem. At each such step, BZ
surface integrals are generated. But each of these integrals vanishes, because
it involves pairwise equal values of the integrand at the opposite points of the
BZ surface. The outer normals to the surface in such points have opposite
orientation and this ensures the cancellation. Note that in this argument both
the periodicity of the derivatives of $n(\vec k)$ and that of $e^{-i \vec k \vec
r}$ are required.  The latter is ensured by $\vec r$ having integer coordinates.

After trasferring all the derivatives one is left with 
\begin{equation}
 r^\alpha \mathcal{C}(\vec{r}) = 
\frac{(-i)^{|\alpha|}}{n_{a}\,(2 \pi)^{m}} \int_{V_{k}} d^{m}k e^{-i \vec{k}
\vec{r}} \partial^\alpha n(\vec{k})\, .
\end{equation}
Since the integrand is again a continuous function, the Riemann-Lebesgue lemma
can be invoked again, ensuring that, indeed, $r^\alpha \mathcal{C}(\vec{r})$
goes to zero for large values of the argument. This concludes the proof. 

The only possibility that the correlation length could diverge is thus
a case where $(2 n_{\sigma} - 1) = \Gamma/\kappa_\sigma$, for which
$n_{\vec{k}} \propto \Delta_{\vec{k}}^{-2}$. For this case, however,
the last term in Eq.~(2b)
in the main text, which reads $(2 n_{\sigma} - 1) \frac{1}{n_a \,
  (2\pi)^{m}} \int_{V_{k}} d^{m}k F_{\vec{k}} n_{\vec{k}}$, diverges
as long as $(2 n_{\sigma} - 1) \ne 0$.  The origin of this divergence
is that $\Delta_{\vec{k}}^{-2}$ at least scales as
$\Delta_{\vec{k}}^{-2} \propto
(k_{\alpha}-\overline{k}_{\alpha})^{-2}$ in the vicinity of a manifold
$\overline{k}$ where $\Delta_{\vec{k}} = 0$ (if $\Delta_{\vec{k}} = 0$
occurs at the boundary of the integration volume the divergence is
even more severe).  We thus conclude that non-exponential decay or a
divergent correlation length can only appear for $\delta = 0$ and $(2
n_{\sigma} - 1) = 0$. Both conditions can only hold for
$\gamma_{a}=0$, i.e. if the photon decay vanishes.

\subsubsection*{Estimates for field correlations in one dimension in
  the limit $N \to \infty$}

\begin{figure*}
  \centering
  \includegraphics[width=14cm]{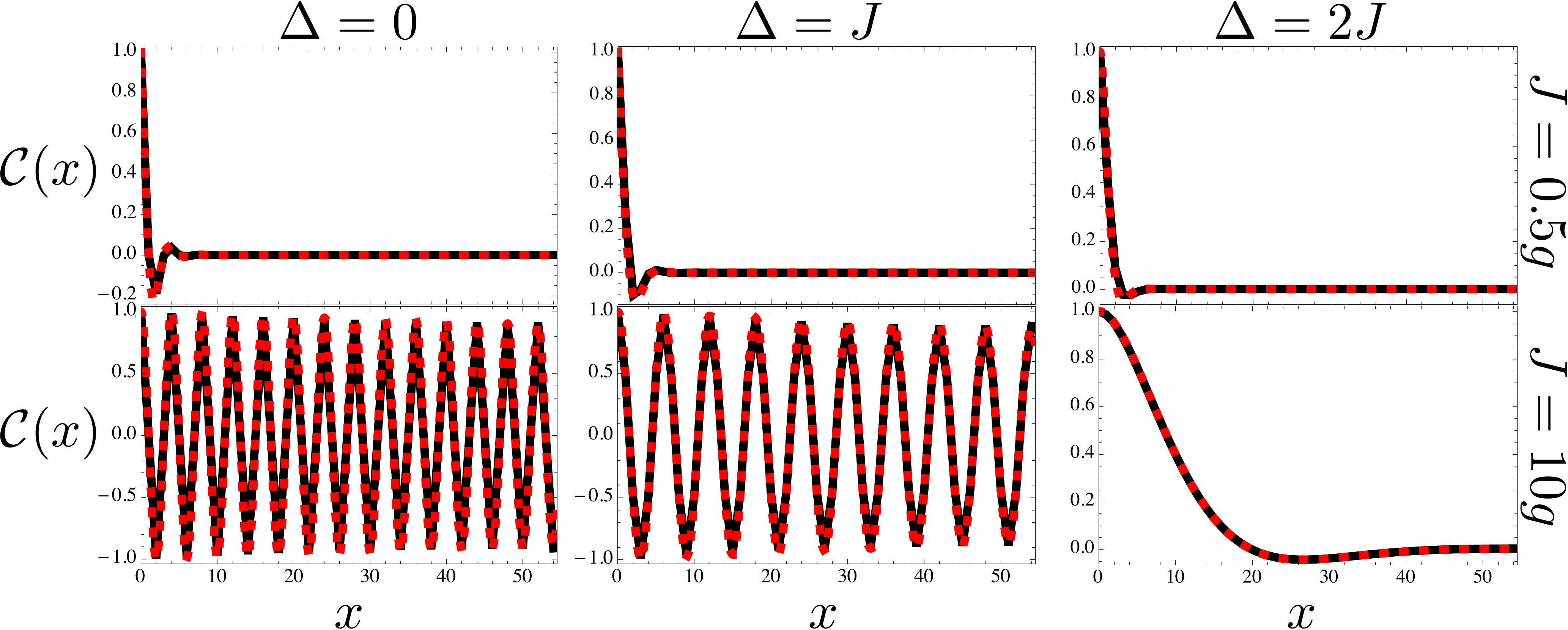}
  \caption{Examples for fits of functions $f(x) = [c_{1} \cos(\nu x) +
    c_{2} \sin(\nu x)] \exp(-\lambda x)$ to the normalized
    correlations $\mathcal{C}(x)$ for $N=108$ and the parameters
    $\Delta$ and $J$ given in the labels of the columns and
    rows. Other parameters are $\gamma_a=0.1g$, $\gamma_\sigma=0.01g$,
    $P_\sigma=5g$.}
  \label{fig:fits}
\end{figure*}

For one dimension, $m=1$, the momentum distribution in the stationary
state reads, $n_{k} = \frac{\kappa_\sigma \Gamma}{4}
\frac{n_{\sigma}}{(\delta/2)^{2} + \Delta_{k}^{2}} $, which is a
Lorentzian in the detunings $\Delta_{k} = \Delta - 2 J \cos k$, and
for $N\to \infty$ the field correlations read,
\begin{equation} 
\mathcal{C}(x) =\frac{1}{n_{a} 2 \pi} \int_{-\pi}^{\pi} dk e^{-ix k} n_k \, . 
\end{equation}
With $n_k$ a real and even function of $k$, it is obvious that $\mathcal{C}(x)$
is also real and even as a function of the distance $x$. Therefore in what
follows we consider only the case $x \geqslant 0$. Up to the prefactor
$\frac{\kappa_\sigma\Gamma n_\sigma}{4 n_a J^2}$, and bearing in mind that 
$x$ takes only integer values, the correlations are obtained by calculating a
Fourier transform of the form
\begin{equation}
C_n = \frac{1}{2\pi}\int_{-\pi}^\pi\, 
\frac{e^{ikn}}{(2\cos k-\widetilde \Delta)^2+\widetilde\delta^2}\, dk, 
\quad n = 0, 1 \dots 
\end{equation}
with the parameters $\widetilde \Delta$ and $\widetilde \delta$ easy to identify
as $\widetilde\Delta = \Delta/J$ and $\widetilde \delta = \delta/(2J)$. One
rearranges the expression under the intergal as
\begin{equation}
\frac{1}{(2\cos k-\widetilde \Delta)^2+\widetilde\delta^2}  =
\frac{1}{2i\widetilde \delta}\, \frac{1}{2\cos k-\widetilde
\Delta-i\widetilde \delta } +\text{c.c.} \, ,
\end{equation}
so that one has to compute
\begin{equation}
C_n = \frac{1}{4\pi i \widetilde \delta}\int_{-\pi}^\pi\, 
\frac{e^{ikn}}{2\cos k- u} \, dk + \text{c.c.} \, , 
\end{equation}
where $u$ denotes the complex quantity $u = \widetilde \Delta+i\widetilde \delta = J^{-1}(\Delta+i\delta/2)$.
This integral is solved by introducing the new variable
$z=e^{ik}$, which runs on the unit circle ${\cal C}_1$,
\begin{equation}
C_n = \frac{-1}{4\pi\delta}\int_{{\cal C}_1} 
\frac{z^n}{z^2-u\,z +1} \, dz + \text{c.c.} \, .
\end{equation}
The poles of the integrand are the roots of the 
denominator $\zeta_{1,2}$, and  satisfy $\zeta_1+\zeta_2 =u$ and 
$\zeta_1\,\zeta_2 = 1$. 
There are two possibilities, either (i) 
$|\zeta_1|<1<|\zeta_2|$, or (ii) $|\zeta_1|=1=|\zeta_2|$. Representing 
the roots as $\zeta_2=e^{\lambda} e^{iq}$ and $\zeta_1=e^{-\lambda}
e^{-iq}$, case (i) amounts to $\lambda>0$ and $\zeta_1$ lying inside the
unit circle. The residue theorem then gives 
\begin{equation}
C_n = 
\frac{i}{2\widetilde \delta}\, 
\frac{1}{\zeta_2-\zeta_1}\, \zeta _1^n + \text{c.c.} \, .
\end{equation}
This shows that the correlations oscillate along the chain with a wave number
$q$ and decay exponentially with the inverse decay length $\lambda$. 

Case (ii) corresponds to $\lambda = 0$, when both roots are found on ${\cal
C}_1$. This takes place when $u = \zeta_1+\zeta_2 = 2\cos q$ i.e.
$u$ is real and belongs to the interval $[-2,2]$. With poles on the integration
path the integral is divergent. Still, it makes sense to consider this as a
limit case, with $u$ approaching the segment $[-2,2]$ of the real axis. Then
$\zeta_1$ approaches the unit circle from within, and the correlation length
$1/\lambda$ goes to infinity. The system becomes critical. The requirements on 
the system parameters for achieving criticality are $\delta \to 0$ and 
$|\Delta| \leqslant 2J $. It also follows that $q$ is the momentum of the 
resonant Bloch mode. 

It is straightforward to relate the quantities $\lambda$ and $q$, to the system
parameters but the expressions are cumbersome. Some qualitative features are 
easily obtained though, and they describe different regimes of correlation 
behaviour. 

A first situation is encountered when $u$ lies in the complex plane far away 
from the critical interval $[-2,2]$. For $\widetilde \Delta $ and $\widetilde \delta$ large,
this corresponds to small $J$-values, since $\widetilde \Delta \propto J^{-1}$ and $\widetilde \delta \propto J^{-1}$
In this case $\lambda$ is large and in the relation $\zeta_1 
+ \zeta_2 = u$ the small root $\zeta_1$ becomes negligible. It follows that 
$\lambda = \ln |\zeta_2| \simeq \ln |u| \propto -\ln J$. 

A completely different behavior is seen when $u$ is close to the segment 
$[-2,2]$. In this regime $J$ is large to make $\widetilde \delta$ small.
Also, $\Delta$, $J$ are of the same magnitude and obey $|\Delta| \leqslant 2J$,
to keep $\widetilde \Delta $ within the limit of the interval. In this case
$\lambda \simeq 0$, both roots are close to the unit circle. Therefore both
contribute to the sum, and one can write
\begin{equation}
\frac{1}{2} u = \frac{1}{2}(\widetilde \Delta +i \widetilde \delta) = \cosh
\lambda \cos q + i \sinh \lambda \sin q \, . 
\label{eq:hyperb_trig}
\end{equation}
With $\lambda$ small, one has $\cosh\lambda \simeq 1$ and $\sinh
\lambda \simeq \lambda$ and by identifying the real and imaginary
parts, it follows that $\cos q = \widetilde \Delta/2 = \Delta /(2J)$ and 
\begin{align}
\lambda & = \frac{\widetilde \delta}{2\sin q}=
\frac{\delta/2}{\sqrt{4J^2-\Delta^2}} \nonumber \\
 & =
\sqrt{\frac{g^2 \Gamma}{\gamma_a(4J^2-\Delta^2)}\left[\frac{\gamma_a
\Gamma}{4g^2}- (2n_\sigma -1)\right] } \, .
\end{align}
With $\Delta$ of the same order as $J$, one obtains $\lambda \propto J^{-1}$.   

The above result holds for $\widetilde \Delta$ not too close to the endpoints
of the critical interval, where $\sin q$ becomes small and division by it gives
rise to large values of $\lambda$. This is seen in the final expression for
$\lambda$, in which $\Delta$ approaching $2J$ leads to a singularity. Therefore
this case requires a separate, more careful consideration, since now $q$ becomes
a small quantity, too.
Expanding up to the second order in terms of the small arguments,
Eq.~(\ref{eq:hyperb_trig}) becomes
\begin{equation}
\frac{1}{2}(\widetilde \Delta +i \widetilde \delta) \simeq 1 + \frac{1}{2}
\lambda^2 - \frac{1}{2} q^2 + i \lambda \,q \, .
\label{eq:second_order}
\end{equation}
To keep the discussion simple we discuss the case $\Delta = 2J$,
or $\widetilde \Delta =2$. Actually this illustrates the more general 
situation in which $1-\widetilde\Delta/2$ is a small quantity of a higher
than second order. Then, from Eq.~(\ref{eq:second_order}) we find
$\lambda=q$ and $\lambda^2 = \widetilde\delta /2= \delta/(4J)$. More precisely
\begin{equation}
\lambda = \left\{ \frac{g^2 \Gamma}{4 \gamma_a J^2}\left[\frac{\gamma_a
\Gamma}{4 g^2} - (2 n_\sigma-1) \right] \right\}^{1/4} \, .
\end{equation}
Note that now $\lambda \propto J^{-1/2}$.


\subsubsection*{Examples for the fits}

In this section we provide some examples for the fits of functions
$f(x) = [c_{1} \cos(\nu x) + c_{2} \sin(\nu x)] \exp(-\lambda x)$ to
the normalized correlations $\mathcal{C}(x)$. These examples are shown
in Fig.~\ref{fig:fits} and illustrate the excellent quality of the
fits.  Only for $J\ll g$ the fitting procedure is more fragile as
correlations decay very fast and are thus indistinguishable from zero
for most values of $x$.

\section{III. Derivation of the emitter spectrum of emission}

In this section we obtain the emitter photoluminescence
spectrum~$S(\Gamma_d,\omega)$, in the lasing regime, where $\Gamma_d$
is the detector linewidth. We make the semiclassical approximation of
substituting the cavity fields by a multimode laser that acts
independently on each of the emitters. That is, we consider the
approximated Hamiltonian~$ H_\mathrm{ML}= \sum_{\vec{r}}[
\omega_\sigma\ud{\sigma_{\vec{r}}} \sigma_{\vec{r}}+
\Omega(t)\ud{\sigma_{\vec{r}}} +\Omega^*(t) \sigma_{\vec{r}}]$, where
$\Omega(t)=\sum_{\vec{k}} g \sqrt{n_{\vec{k}}/N} e^{-i\omega_{\vec{k}}
  t}$ is the time-dependent multimode field. Additionally, the
emitters are still being excited by the incoherent pump and decay that
act on their dynamics through the usual Lindblad forms. There is no
steady state for this approximated model (for $N>1$) but a
quasi-steady state, that is, an ever oscillating solution for the
density matrix elements around a mean point. Such mean point is given
(approximately) by the exact solution of the full master equation or
the rate equations, which do have a steady state. That is,
$\sum_{\vec{k}}
G_{\vec{k}\vec{r}}\mean{p_{\vec{k}}\sigma_{\vec{r}}^\dagger}e^{-i\omega_{\vec{k}}
  t}$ is well estimated by $\Omega(t)
\mean{\sigma_{\vec{r}}^\dagger}_\mathrm{ML}$, where
$\mean{\cdot}_{ML}$ is the mean value obtained with the approximated
master equation and Hamiltonian $H_{ML}$ for the emitters only. The
fact that the first term is $\vec{r}$-independent, compels $\Omega(t)$
to be $\vec{r}$-independent as well. We describe the resulting
time-dependent dynamics in the following way: First, we solve the new
master equation with $H_\mathrm{ML}$, and obtain its time-dependent
spectrum of emission~\cite{eberly77a,eberly80b},
$S_{\mathrm{ML}}(\Gamma_d,\omega,t)$, by coupling the emitter very
weakly to another two-level system, which radiatively decays at a
rate~$\Gamma_d$, and plays the role of the detector. The population of
this detector is exactly the time-dependent spectrum of our
emitter~\cite{delvalle12a}. Then, we take its average over time, once
the quasi-steady state is reached, starting at a point in time which
we call~$t_0$: $S(\Gamma_d,\omega)\approx \int_{t_0}^{t_0+T}
S_{\mathrm{ML}}(\Gamma_d,\omega,t)dt /T$. This is a very good
approximation in the case $N=1$~\cite{delvalle10d,delvalle11a} for
which there is a simple analytical
formula~\cite{delvalle13c_mathematica}. The Rayleigh peak, produced by
the elastically scattered cavity laser field, is pinned at the cavity
frequency, $\omega=\omega_a$, and has a small linewidth given by the
detector only~$\Gamma_d$ (as in this approximation the cavity has an
infinitely long lifetime). We used $\Gamma_d=0.3g$ to plot the spectra
in Fig.~3(j)--(l) of the main text.

\end{document}